\documentclass[12pt]{article}
\usepackage{setspace}
\usepackage{subfigure}
\usepackage{graphicx}
\usepackage{caption}
\usepackage{amssymb}
\usepackage{cite}
\usepackage{color}
\usepackage{booktabs}
\usepackage{tabularx}
\usepackage{array}
\usepackage{makecell}
\usepackage{amsmath}
\usepackage{xr}
\usepackage{upgreek}
\usepackage{geometry}
\usepackage{siunitx}
\usepackage{graphicx}
\usepackage{subfigure}

\usepackage[pagestyles]{titlesec}
\usepackage{xurl}  
\usepackage{sidecap}  
\usepackage{textcomp}  
\usepackage{xr-hyper} 
\usepackage{hyperref}
\hypersetup{
	colorlinks=true,
	linkcolor=cyan,
	filecolor=blue,
	urlcolor=blue,
	citecolor=green,
}

\newcommand{\PreserveBackslash}[1]{\let\temp=\\#1\let\\=\temp}
\newcolumntype{C}[1]{>{\PreserveBackslash\centering}p{#1}}
\newcolumntype{R}[1]{>{\PreserveBackslash\raggedleft}p{#1}}
\newcolumntype{L}[1]{>{\PreserveBackslash\raggedright}p{#1}}
\usepackage[labelfont=bf,labelsep=period,justification=raggedright, font=small]{caption}

\begin{document}
\thispagestyle{empty}
\begin{flushleft}
{\Large

	\textbf{PepGB: Facilitating peptide early drug discovery via graph neural networks}
}
\\
\vspace{5pt}
Yipin Lei$^{1 \sharp}$,
Xu Wang$^{2 \sharp}$,
Meng Fang$^{1}$,
Han Li$^{1}$,
Xiang Li$^{4}$,
Jianyang Zeng$^{3,\ast}$
\\
\vspace{5pt}
\bf{1} Institute for Interdisciplinary Information Sciences, Tsinghua University, Beijing 100084, China.\\
\bf{2} Machine Learning Department, Silexon AI Technology Co., Ltd., Nanjing 210046, China.\\
\bf{3} School of Engineering, Westlake University, Zhejiang Province, Hangzhou 310030, China. \\
\bf{4} School of Pharmacy, Second Military Medical University, Shanghai 200433, China. \\

$\sharp$ These authors contributed equally. \\
$\ast$ All correspondence should be addressed to zengjy@westlake.edu.cn.
\clearpage
\end{flushleft}

\section*{Abstract}
\normalsize
Peptides offer great biomedical potential and serve as promising drug candidates. Currently, the majority of approved peptide drugs are directly derived from well-explored natural human peptides. It is quite necessary to utilize advanced deep learning techniques to identify novel peptide drugs in the vast, unexplored biochemical space. Despite various in silico methods having been developed to accelerate peptide early drug discovery, existing models face challenges of overfitting and lacking generalizability due to the limited size, imbalanced distribution and inconsistent quality of experimental data. In this study, we propose PepGB, a deep learning framework to facilitate peptide early drug discovery by predicting peptide-protein interactions (PepPIs). Employing graph neural networks, PepGB incorporates a fine-grained perturbation module and a dual-view objective with contrastive learning-based peptide pre-trained representation to predict PepPIs. Through rigorous evaluations, we demonstrated that PepGB greatly outperforms baselines and can accurately identify PepPIs for novel targets and peptide hits, thereby contributing to the target identification and hit discovery processes. Next, we derive an extended version, diPepGB, to tackle the bottleneck of modeling highly imbalanced data prevalent in lead generation and optimization processes. Utilizing directed edges to represent relative binding strength between two peptide nodes, diPepGB achieves superior performance in real-world assays. In summary, our proposed frameworks can serve as potent tools to facilitate peptide early drug discovery.

\section{Introduction}
Peptides, such as hormones, signal peptides and neuropeptides, play pivotal roles in various fundamental cellular functions through interacting with proteins and other molecules \cite{pep_th1,pep_th2}. For instance, peptides can modulate pathogenic protein-protein interactions by binding to one of the proteins as well as form interactions along flat and hydrophobic interfaces of the ``undruggable'' proteins where conventional small-molecules are not suited \cite{pep_th2}. Consequently, hundreds of peptide therapeutics, such as Semaglutide and Liraglutide, are approved or currently under-evaluated in clinical trials \cite{pep_drug}. Since these low-hanging fruits have already been picked, there arises an imperative need to explore new paths beyond traditional approaches for peptide drug discovery. 

The drug discovery process is a long, costly, and high-risk journey that can take up to 15 years \cite{pep_drug}. As the starting point, early drug discovery is a critical phase in the drug discovery process as it lays the foundation for the development of effective and safe drug candidates. It typically involves target identification, hit discovery, hit-to-lead, lead generation and optimization, \textit{in vivo} and \textit{in vitro} assays \cite{dd}. Developing computational tools to identify novel peptide-protein interactions (PepPIs) in the unknown broad biochemical space can largely improve the overall efficiency and success rates of peptide early drug discovery. For instance, researchers have developed several peptide docking tools, such as GalaxyPepDock \cite{galaxy}, MDockPeP \cite{mdockpep} and HPEPDOCK \cite{hpepdock} to generate potential complex structures through molecule dynamics and energy optimization. Additionally, various sequence-based deep learning approaches have been developed to predict interactions involving proteins and diverse ligands, e.g., protein-protein interactions \cite{d_script,tt}, compound-protein interactions \cite{monn,deepdta} and protein-DNA/RNA interactions \cite{prot_rna,bgfe}. Our previous work CAMP \cite{camp} is the first deep learning framework to predict general PepPIs. Besides, there also exist several deep learning methods to identify the peptide binding sites of the proteins \cite{pepnn} or utilizing generative models to design proteins capable of binding to peptides \cite{baker_pep}.

\begin{figure}[htbp]
	\centering
	\includegraphics[width=1\textwidth]{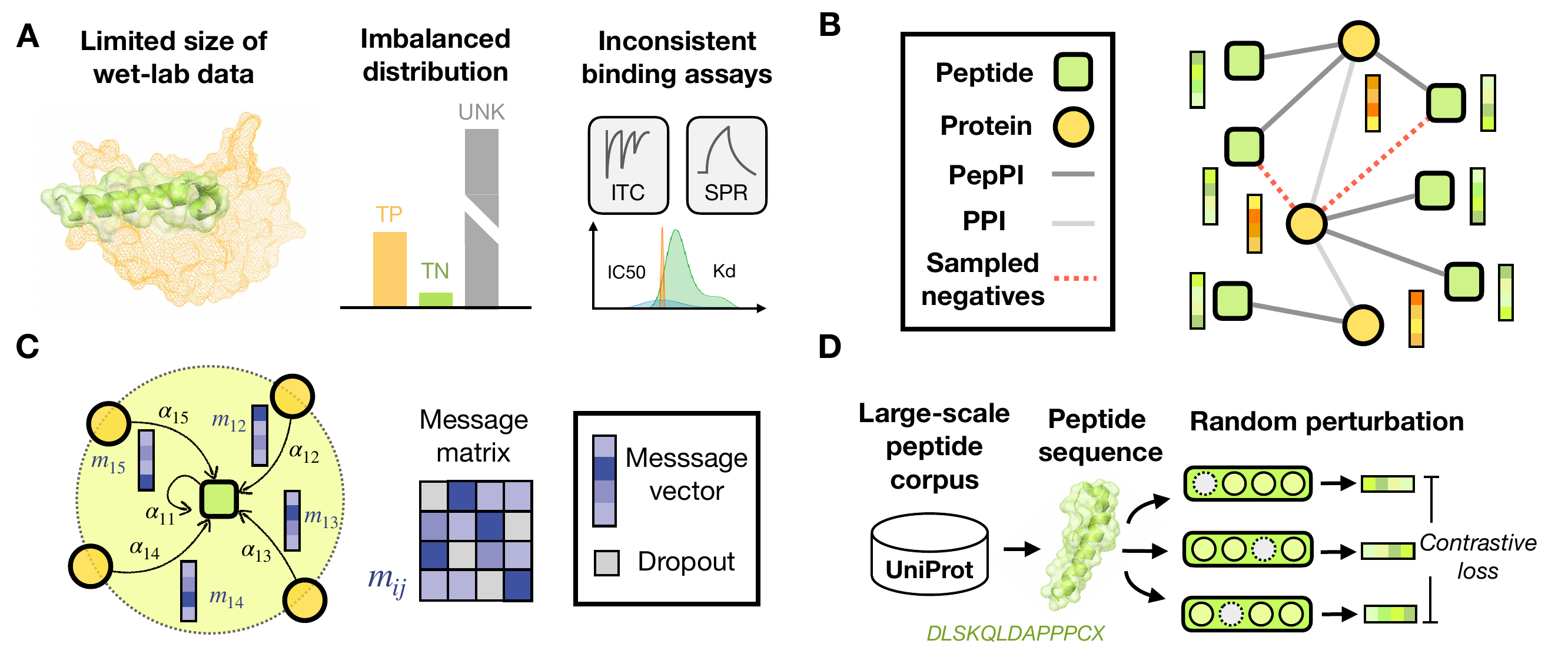}
	\caption{
    	Overview of PepGB. \textbf{A} The motivation of our proposed framework is to address empirical challenges in drug discovery, i.e., we only have limited experimental interaction data; popular targets or peptides are more frequently measured and the remaining unknown interactions do not always represent negatives; binding labels are inconsistent due to batch effects and systematic errors. \textbf{B} PepGB is a heterogeneous graph-based framework to predict PepPIs. Pre-trained sequence embeddings are served as node features. Protein-protein interactions of existing protein nodes are supplemented for additional message passing. \textbf{C} PepGB exploits graph attention neural network (GAT) to update node features via aggregation from neighboring nodes. To avoid overfitting and improve generalizability, the DropMessage module randomly applies dropout on each element of the message passing matrix. \textbf{D} The tailor-made contrastive learning-based pre-training strategy aims to learn peptide representation from a large-scale peptide sequence database.
     }
    
	\label{fig:fig1}
\end{figure}

However, docking approaches are time-consuming and less effective for high-throughput virtual screening. Furthermore, these structure-based methods, even including the advanced AlphaFold-multimers \cite{af_mul} and other 3D geometric models \cite{pronet,gearnet}, face a critical challenge that many peptides tend to be partially unstructured in isolation \cite{baker_pep} and may exhibit huge conformation changes upon binding to the target protein. The inherent flexibility largely hinders us from systematically modeling the binding activities from structural perspectives. Another challenge is the lack of generalizability when applying existing deep models in early drug discovery. Despite intensive efforts made to improve model performance on public benchmarks via traditional cross-validation or random-split test sets, we witness poor performance when applying them on dissimilar data compared to their training sets \cite{camp,monn,teim}. This discrepancy becomes particularly evident when predicting interactions for novel targets or ligands, which are common scenarios in peptide drug discovery. As shown in Fig. \ref{fig:fig1}A, the reasons for this are threefold. Firstly, the scarcity of interaction data, limited by the expensive and time-consuming data generation process in wet-lab. Thus it is quite easy for deep models to overfit the limited training data if we do not carefully design and evaluate the model. Secondly, the imbalanced nature of interaction data, influenced by the ``exposure bias'' \cite{bias}, which means that only a small portion of peptides and proteins are studied so that unobserved interactions do not always represent true negatives. Lastly, the dependence of biological labels (e.g., binding affinities) on experimental conditions and protocols, requiring the prediction model to possess robust generalizability to bear such uncertainty and inconsistency.

In this study, we propose PepGB (\textbf{Pe}tide-\textbf{p}rotein interaction via \textbf{G}raph neural network for \textbf{B}inary prediction), a heterogeneous graph-based deep learning framework for predicting peptide-protein interactions, to facilitate peptide early drug discovery. PepGB exploits a graph attention neural network to capture the topological information among a limited number of peptides and proteins. To alleviate the problem of overfitting, PepGB is equipped with a fine-grained perturbation during message passing process. We also incorporate a dual-view loss to prevent our model from the influence of imbalanced and uncertain negatives. We carefully investigated PepGB and other state-of-the-art methods under strict evaluation settings and demonstrated that PepGB possesses good generalizability to identify novel targets and peptide hits.

To address the challenge of modeling highly imbalanced data that are prevalent in lead generation and optimization processes, we derive an extended version, a \textbf{di}rected graph-based framework called diPepGB. Taking into account possible experimental errors, we elaborately designed a rank-based strategy to construct error-tolerated directed edges. Evaluation results illustrated that diPepGB alleviates the issue of highly uneven topology and successfully characterizes valuable peptide leads.

In summary, our key contributions lie in formulating peptide-protein interaction prediction as link prediction using graph neural networks and applying PepGB and diPepGB to key stages in drug discovery. Comprehensive evaluation shows that the graph-based paradigm significantly enhances model performance, highlighting it capacity as a powerful tool in peptide early drug discovery. We anticipate that our study can provide insightful perspectives and benefit future designs of peptide therapeutics.

\section{Methods}

\subsection{Problem formulation}\label{sec:PepPI_graph}
In this work, we harness the powerful graph structure to model peptide-protein interactions (PepPIs) and leverage the effective graph neural network to learn the intricate interactions between peptides and proteins.

\textbf{PepGB} We first construct a heterogeneous graph to represent the PepPIs. Let $G=(V,E)$ denote a graph, where $V = \{v_i |i=1,..,n\}$ represent the set of $n$ nodes and $E = \{e_j |j=1,...,m\}$ represent the set of $m$ edges. The heterogeneous graph also incorporates a node type mapping function $\phi : V \to O_v$ and a edge type mapping function $\psi : E \to R_e$. As shown in Fig. \ref{fig:fig1}B, the node type set $O_v$ contains peptide and protein nodes, and the edge type set $R_e$ includes peptide-protein interaction (PepPI) edges and protein-protein interaction (PPI) edges. It is noteworthy that the PepPI edges are considered as our prime edges since we focus on predicting PepPIs, and we additionally incorporate known PPI edges to the PepPI graph to enable the sharing of the PepPI binding patterns between proteins involved in PPIs through the message passing process. In such a manner, the occurrence of PepPI can be formulated as a link prediction task on PepPI edges.

\begin{figure}[htbp]
	\centering
	\includegraphics[width=1\textwidth]{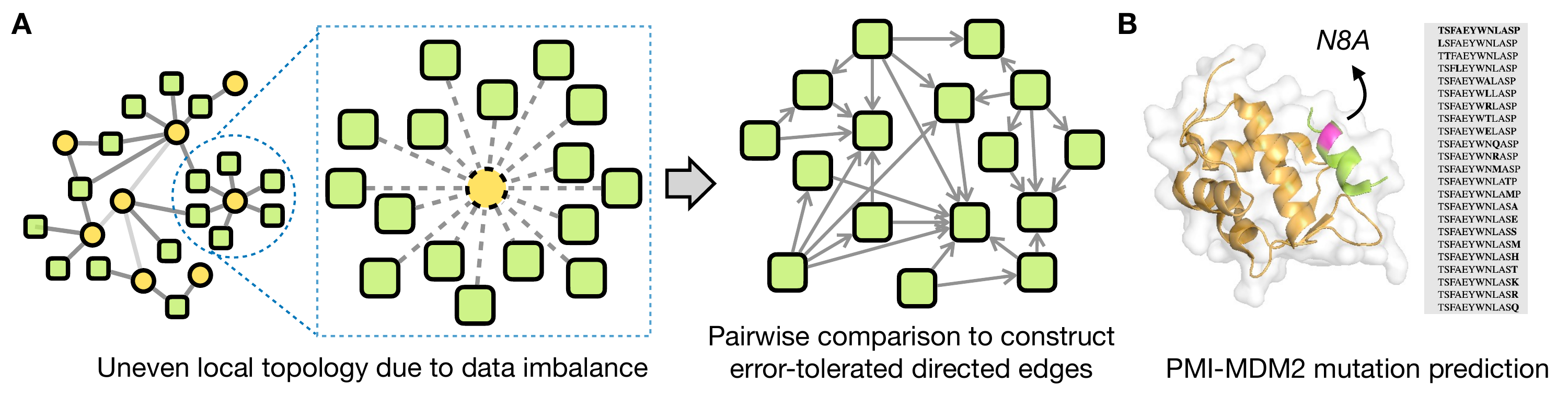}
	\caption{
        An illustration of the framework of diPepGB. \textbf{A} diPepGB aims to address the imbalance nature of experimental data, which would lead to the uneven local topology on PepPI graph. The ``firework'' style sub-graph can be formulated into a directed graph by constructing pairwise comparison of the peptide binding strength. To make diPepGB maintain robust against systematic errors, we define error-tolerated directed edges sourcing from peptide nodes with significantly stronger affinities and pointing to the peptide nodes with weaker affinities. \textbf{B} We illustrated the application of diPepGB in lead optimization using binding assays of peptide analogs binding to the oncoprotein MDM2.
     }
	\label{fig:fig2}
\end{figure}

\textbf{diPepGB}
In lead generation and optimization processes, researchers refine the initial hits by affinity selection, mutation analysis and target-focused libraries to obtain promising leads, where the binding affinities of a series of peptide analogs targeting the same protein are measured, resulting in the prevalence of extremely imbalanced data in related assays. This consequently introduces a notable challenge: for such highly imbalanced experimental data, the local topology of the PepPI graph becomes quite uneven that resembles a ``firework'' (Fig. \ref{fig:fig2}A). From a theoretical perspective, graph neural networks may exhibit suboptimal performance since the message passing process is less effective due to the lack of directly mutual edges between the peptide nodes. To tackle this limitation, we derive an extended framework called diPepGB, utilizing a directed graph to profile the affinity variation between peptide mutants from individual assays. As shown in Fig. \ref{fig:fig2}A, the directed edge is defined as sourcing from the peptides with significantly stronger affinities and pointing to those with weaker affinities. Due to different wet-lab conditions and experimental protocols, there often exist systematic errors in real-world assays \cite{error}. For instance, a peptide with an affinity of 10 $nM$ may roughly exhibit a similar binding strength to a peptide with an affinity of 7 $nM$ measured under the same experimental conditions. To address this, peptides from the same assay are categorized as ``stronge'' and ``weaker'' only if one binds to the target protein at least threefold stronger than the other. This strategy allows us to establish a directed graph to profile the imbalanced data. Specifically,  we construct a homogeneous graph denoted as $G=(V,E)$, where $V = \{v_i |i=1,..,n\}$ represents the set of $n$ peptide nodes and $E = \{e_{jk}|j\in V, k\in V \}$ represents the set of $m$ directed edges sourcing from peptide node $v_j$ and pointing to peptide node $v_k$ from the same assay. In the directed version, we exclude the original edges between proteins and binding peptides since the original definition conflicts with the rank-based formulation. Notably, peptides from different assays may occur in the same graph without any edges connecting them. In this way, the task can be formulated as predicting directed links between peptide nodes.

\subsection{Datasets}\label{sec:Dataset}

In this study, we define peptides as containing no more than 50 residues and proteins have no length constraints.  All non-standard amino acids are replaced by ``X''. We curated three datasets in our experiments, including a binary interaction benchmark dataset comprising both PepPIs and complementary PPIs to train PepGB, a mutation dataset featuring peptide analogs binding to the same oncogenic protein MDM2 from three mutation assays to validate diPepGB, and a large-scale peptide sequence dataset for pre-training. Additional information about the datasets is available in Supplementary S1.

\begin{itemize}
\item[$\bullet$] \textbf{Binary interaction benchmark} We constructed a binary PepPI benchmark dataset using protein–peptide complex structures sourced from the RCSB PDB \cite{pdb1}. Following our previous workflow \cite{camp}, we identified 7,233 prime PepPI edges covering 5,283 protein nodes and 3,318 peptide nodes. To create negatives, we randomly shuffled non-interacting pairs of peptides and proteins, thereby generating peptide-protein edges absent in the original PepPI graph. To facilitate the PepPI graph, we incorporated known PPI data involving the existing protein nodes in the graph. Here we exclusively considered protein-protein interactions through physical contacts. Our PPI data were derived from two sources: 191 mapped PPI complexes data from \cite{masif} and 1,737 PPI data with positive experimental scores or identified as physical bindings in the BIOGRID database \cite{biogrid1}.

\item[$\bullet$] \textbf{Peptide mutation dataset}
The oncoprotein MDM2, a crucial target for anticancer therapy, negatively regulates the bioactivity of the tumor suppressor protein p53 \cite{pmi1}. Previous work \cite{pmi2} identified PMI, a potent 12-mer peptide inhibitor for MDM2, exhibiting low nanomolar affinity. Multiple mutational analysis, including alanine scanning, were conducted to identify crucial residues of PMI peptide for lead optimization. Here, we collected 100 PMI peptide analogs from \cite{pmi3}, 12 mutants derived from a single-position alanine scanning assay from \cite{pmi2} and 9 mutants derived from a two-position alanine scanning assay from \cite{pmi4}. For each assay, we performed pairwise comparisons to construct directed edges pointing from the significant strong binders to the weak binders based on binding affinities. To alleviate systematic errors, a strong binder is defined as exhibiting a binding affinity at least threefold stronger than the weak binder.

\item[$\bullet$] \textbf{Peptide corpus for pre-training} We extracted peptides with sequence length within 50 from UniProt \cite{uniprot1} to construct a large-scale peptide pre-training dataset. In total, we obtained 3,917,987 peptide sequences for pre-training. All non-standard amino acids were replaced with ``X''.
\end{itemize}

\subsection{Model architecture}\label{sec:Graph_archi}
\subsubsection{Graph attention network}
Graph neural networks (GNNs) are powerful tools as they can preserve rich structural information through conducting message passing across nodes in graphs. As shown in Fig. \ref{fig:fig1}C, we exploit the Graph Attention Network (GAT) \cite{gat} as our backbone architecture due to its proven expressiveness on massive benchmarks \cite{graph}. 

Formally, each node $v_i$ in the $t+1$-th layer in the interaction graph updates its feature by propagating the messages from its neighbors 

\begin{equation}
h_{i}^{(t+1)} = \gamma(W^{(t+1)}\cdot [\sum_{j \in N_i} \alpha_{ij}^{(t)}h_j^{(t)}+\alpha_{ii}^{(t)}h_i^{(t)}]),
\end{equation}\label{eq:agg}

\begin{equation}
\alpha_{ij}^{(t)} = \frac{A^{(t)}(h_i^{(t)},h_j^{(t)})}{\sum_{k \in N_i}A^{(t)}(h_i^{(t)},h_{k}^{(t)})}, \label{eq:att}
\end{equation}

where $h_i^{(t)}$ and $h_j^{(t)}$ stand for the node features of $v_i$ and $v_j$ in the $t$-th layer, respectively. $N_i$ is the first-order neighborhood of node $i$ in the graph, $\gamma$ denotes a learnable function, $W^{(t)}$ is a weight matrix and $\alpha_{ij}$ stands for the attention coefficient indicating the importance of node $j$ to node $i$. $\alpha_{ij}$ is calculated by an attention mechanism $A^{(k)}$ \cite{attention}.


\subsubsection{DropMessage module}
To alleviate the problem of overfitting, we employ a fine-grained perturbation module, called DropMessage \cite{dropmessage}, to perform a random dropping operation on the message matrix during the message passing process. More specifically, the message passing part in the Eq. \ref{eq:agg} can also be denoted as a message matrix $M$:
 \begin{equation}
\mathbf{M}_{i,j}^{(t)} = \phi^{(t)}(h_i^{(t)},h_j^{(t)},e_{ij}), 
\vspace{-5pt}
\end{equation}\label{eq:message}
where $\phi^{(t)}$ is a differentiable function and $e_{ij}$ stands for the edge feature between node $i$ and node $j$.

Essentially, DropMessage conducts randomly masking on elements from the message matrix $M$ with the dropping rate $p$, which indicates that in total $p|\mathbf{M}|$ elements will be masked in expectation (Fig. \ref{fig:fig1}C). For each element $m_{ij} \in \mathbf{M}$, we generate an independent mask according to a Bernoulli distribution $Bernolli(1-p)$. Then we scale the masked matrix by $\frac{1}{(1-p)}$ to guarantee the perturbed message matrix to have the same expectation as the original one. Such strategy has proven to theoretically preserve information diversity and practically improve the generalization ability of GNN models \cite{dropmessage}.

\subsubsection{Dual-view objectives}
Apart from optimizing the standard binary cross-entropy loss \cite{bce}, we also incorporate an AUC min-max-margin loss to directly maximize the training AUC score (i.e., the area under the ROC curve), which has proven to be robust to noisy and easy data \cite{auc}. Formally, the AUC min-max margin loss is defined as 
 \begin{equation}
 \begin{split}
L_{AUC}(\mathbf{w}) = \mathbb{E}[(g_{\mathbf{W}}(x)-\alpha(\mathbf{w}))^2| y=1] 
\\
+ \mathbb{E}[(g_{\mathbf{W}}(x\prime)-b(\mathbf{w}))^2| y\prime=1] 
\\
+ \mathop{max}\limits_{\alpha \ge0}2\alpha(m-a(\mathbf{w})+b(\mathbf{w})))-\alpha^2, 
\end{split}
\end{equation}\label{eq:auc_loss}

where $(x,y)$ stands for a positive pair and $(x\prime,y\prime)$ stands for a negative pair, $g_{\mathbf{W}}(x)$ is the graph output given the GNN parameters $\mathbf{W}$, $a(\mathbf{w})=\mathbb{E}[g_{\mathbf{W}}(\mathbf{x})|y=1]$, $b(\mathbf{w})=\mathbb{E}[g_{\mathbf{W}}(\mathbf{x\prime})|y\prime=1]$, $\alpha$ is a non-negative constraint and $m$ is a hyper-parameter that defines the desired margin between $a(\mathbf{w})$ and $b(\mathbf{w})$. Then the overall loss of the link prediction for PepGB is illustrated as follows:
 \begin{equation}
L = \eta L_{BCE} + (1-\eta) L_{AUC}
\end{equation}\label{eq:auc_loss}
where $L_{BCE}$ is the standard binary cross-entropy loss and $\eta$ is a hyper-parameter.

There are several benefits of choosing the AUC min-max-margin loss. First, biological data are usually imbalanced, with the number of experimentally measured positives is usually much less than the negatives. AUC essentially measures how well the model ranks true positives higher than negatives and thus can remain robust against such imbalanced data. Second, for our PepPI graph, the true negative edges are wrapped in plenty of ``unknown'' edges due to the ``exposure bias''. During the training process, PepGB randomly samples negative edges from those ``unknown'' edges, which may actually contain positives but have not be identified yet. A previous study \cite{auc} has theoretically proven that AUC min-max-margin loss can alleviate the sensitivity to such level of noisy data. 

\subsection{Contrastive learning-based peptide pre-training}\label{sec:CL_encoder}
Since acquiring labelled interaction data is quite time-consuming and expensive in wet-lab, it is a huge challenge for deep models to generalize well on the broad biochemical space with limited training data. To mitigate this challenge, we develop a self-supervised learning framework that leverages large-scale unlabelled data ($\sim$3.9 million peptide sequences) for peptide pre-training. Specifically, we augment the input peptide sequence by introducing random noise to latent vectors and apply a contrastive estimator to maximize the consistency of augmentations of the same peptide and minimize that of different peptides (Fig. \ref{fig:fig1}D). 

While there exist alternative augmentation strategies like masking individual residues (token-level) and subsampling k-mer (subsequence-level), for peptides with relatively short sequences, augmentation at such levels may conflict with biological realities that even a mutation at a single position can result in significant drops in binding activities, known as activity cliffs. Therefore, we exclusively adopt an embedding-level dropout operation, which can be considered as the minimal augmentation on hidden representations. The previous study \cite{simcse} has shown that this strikingly simple approach outperforms other methods and contributes to improve model performance.

More specifically, we first use a pre-trained protein language model ESM \cite{esm} to encode the amino acid sequences as the initialized feature $h_i = f_{\theta}(x_i)$, where $x_i$ is the input peptide sequence. Then we simply apply the standard dropout twice to acquire a ``positive'' pair and other peptides from the same batch are considered as ``negatives'', and our pre-training objective is to characterize the positive peptide pairs among the negative pairs. Here, we follow the contrastive framework in \cite{simcse} and take the InfoNCE loss \cite{infonce}:
 \begin{equation}
l_i = -log\frac{exp(sim(h_{i},h_{i}^{+})/\tau)}{\sum_{k=1}^{N}exp(sim(h_i,h_j)/\tau)},
\end{equation}\label{eq:infonce_loss}
where $(h_i,h_{i}^{+})$ denotes the augmented embeddings of the same peptide, $\tau$ is a temperature hyperparameter, $N$ is the batch size and the cosine similarity between two embeddings is $sim(h_i,h_j) = \frac{h_i^T\cdot h_j}{\parallel h_i \parallel \cdot \parallel h_j \parallel}$.

Then, for each peptide node $v_i$, our pre-trained encoder generates a feature matrix $h_i^{0}\in \mathbb{R}^{D \times L}$, where $D$ is the dimension of hidden states and $L$ is the peptide sequence length. An average pooling layer is additionally applied to get the final node embedding $h_i\in \mathbb{R}^{D}$.

\subsection{Experimental setups}
\subsubsection{Baseline methods}\
We compared PepGB with six baselines, including CAMP \cite{camp}, CAMP-esm, D-script \cite{d_script}, Topsy-Turvy \cite{tt}, DeepDTA-seq \cite{deepdta} and Transformer \cite{attention}. CAMP is our previous work that predicts binary PepPIs via convolutional neural network and self-attention mechanism. The model leverages pre-processed sequence-based features of a peptide-protein pair to generate an interaction score. CAMP-esm is its variation, which utilizes pre-trained features from a protein language model ESM \cite{esm} as input. To our best knowledge, these two baselines are the only deep learning methods particularly designed to predict binary PepPIs. We also tried PepNN \cite{pepnn}, a deep learning approach aiming to identify peptide binding residues on the protein surface using peptide and protein sequence as input. However, the AUC score oscillated around 0.5, suggesting that the framework might not be suitable for transferring to predict PepPIs (implementation details are available in Supplementary S3.1). In addition, we adopted two sequence-based deep learning frameworks designed for predicting protein-protein interactions (PPIs), i.e., D-script and Topsy-Turvy. D-script employs a pre-trained protein language model \cite{d_script} as a feature extractor and predicts physical PPIs via inter-protein contact maps. Topsy-Turvy, another sequence-based framework, exploits a transfer learning strategy to learn both global and molecular-level PPIs. DeepDTA-seq, a deep learning method designed for predicting binding affinities of drug-target interactions, was also included. Furthermore, we incorporated a Transformer-based model as a reference. As a popular attention-based architecture \cite{attention}, Transformer is widely applied to model biological sequences. For all baselines, peptides and proteins without known evidence of interactions were randomly paired to generate negatives, maintaining a positive-negative ratio of 1:5. Additional details regarding the baselines can be found in Supplementary S3.1.

\begin{figure}[htbp]
	\centering
	\includegraphics[width=1\textwidth]{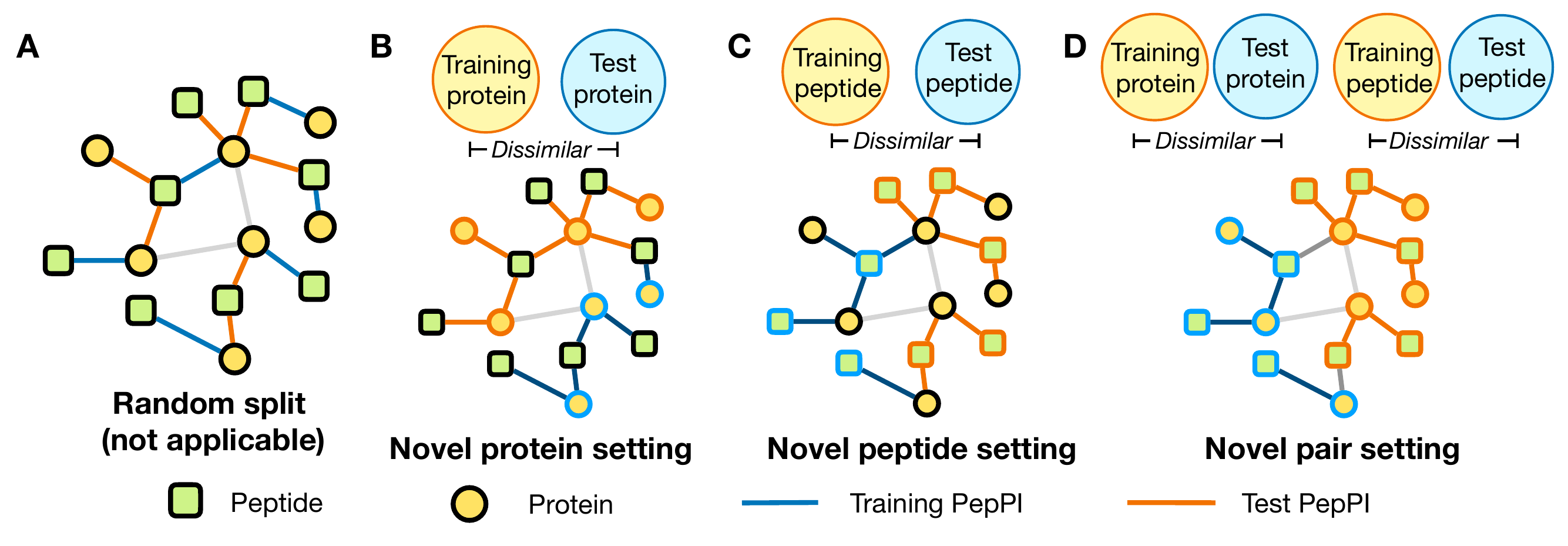}
	\caption{
    	The validation settings to evaluate PepGB. \textbf{A} Computational models commonly adopt the random-split setting for performance evaluation, leading to over-optimistic results. To mimic more realistic scenarios, nodes are first clustered into groups and then different groups are assigned to the training and test sets, respectively. \textbf{B} The ``novel protein setting'' guarantees the training edges and test edges do not share similar protein nodes. \textbf{C} The ``novel peptide setting'' guarantees the training edges and test edges do not share similar peptide nodes. \textbf{D} The ``novel pair setting'' guarantees the training edges and test edges neither share similar peptide nodes nor share similar protein nodes.The validation settings to evaluate PepGB. \textbf{A} Computational models commonly adopt the random-split setting for performance evaluation, leading to over-optimistic results. To mimic more realistic scenarios, nodes are first clustered into groups and then different groups are assigned to the training and test sets, respectively. \textbf{B} The ``novel protein setting'' guarantees the training edges and test edges do not share similar protein nodes. \textbf{C} The ``novel peptide setting'' guarantees the training edges and test edges do not share similar peptide nodes. \textbf{D} The ``novel pair setting'' guarantees the training edges and test edges neither share similar peptide nodes nor share similar protein nodes.
     }
	\label{fig:fig3}
\end{figure}

\subsubsection{Validation settings}
Biological data usually contain ``redundant'' interaction data (e.g., one protein may have many similar binding analogs or vice versa). Given that many evaluation protocols neglect this issue, previous models may demonstrate over-optimistic performance and risk a lack of generalizability (Fig. \ref{fig:fig3}A). Specifically, the presence of similar peptides or proteins in both the training and validation sets, referred to as``trivial cases", may mislead the model performance. 

To rigorously evaluate the model, we adopted a cluster-based cross-validation strategy that has proven effective in avoiding the impact of redundant data \cite{camp,monn,teim}. The strategy ensures the absence of shared similar sequences between training and validation datasets, and thus simulates a more realistic scenario for target and hit identification. The ``novel peptide setting'' and ``novel protein setting'' were evaluated through five-fold cross-validations while the ``novel pair setting'' was through a nine-fold cross-validation (additional details can be found in Supplementary S2.1).

Therefore, we evaluated the model under three different settings: the ``novel peptide setting'' simulating a situation where we aim to identify possible targets for a novel peptide (no similar peptides are shared across the training and validation dataset, Fig. \ref{fig:fig3}B), the ``novel protein setting'' simulating a situation where we aim to screen peptide hits for a novel protein (no similar proteins are shared across the training and validation dataset, Fig. \ref{fig:fig3}C) and the ``novel pair setting'' simulating a situation where we aim to characterize binding activities between novel peptide candidates and proteins (neither similar peptides nor similar proteins are shared across the training and validation dataset, Fig. \ref{fig:fig3}D). Then we conducted  cross-validations based on the clustered data for model evaluation (further details can be found in Supplementary S2.2). 

The training details and hyper-parameters configuration can be found in Supplementary S4.

\section{Results}

\subsection{PepGB greatly outperforms baselines in PepPI prediction for novel targets and peptide hits}\label{sec:5_cv}
\begin{figure}[htbp]
	\centering
	\includegraphics[width=1\textwidth]{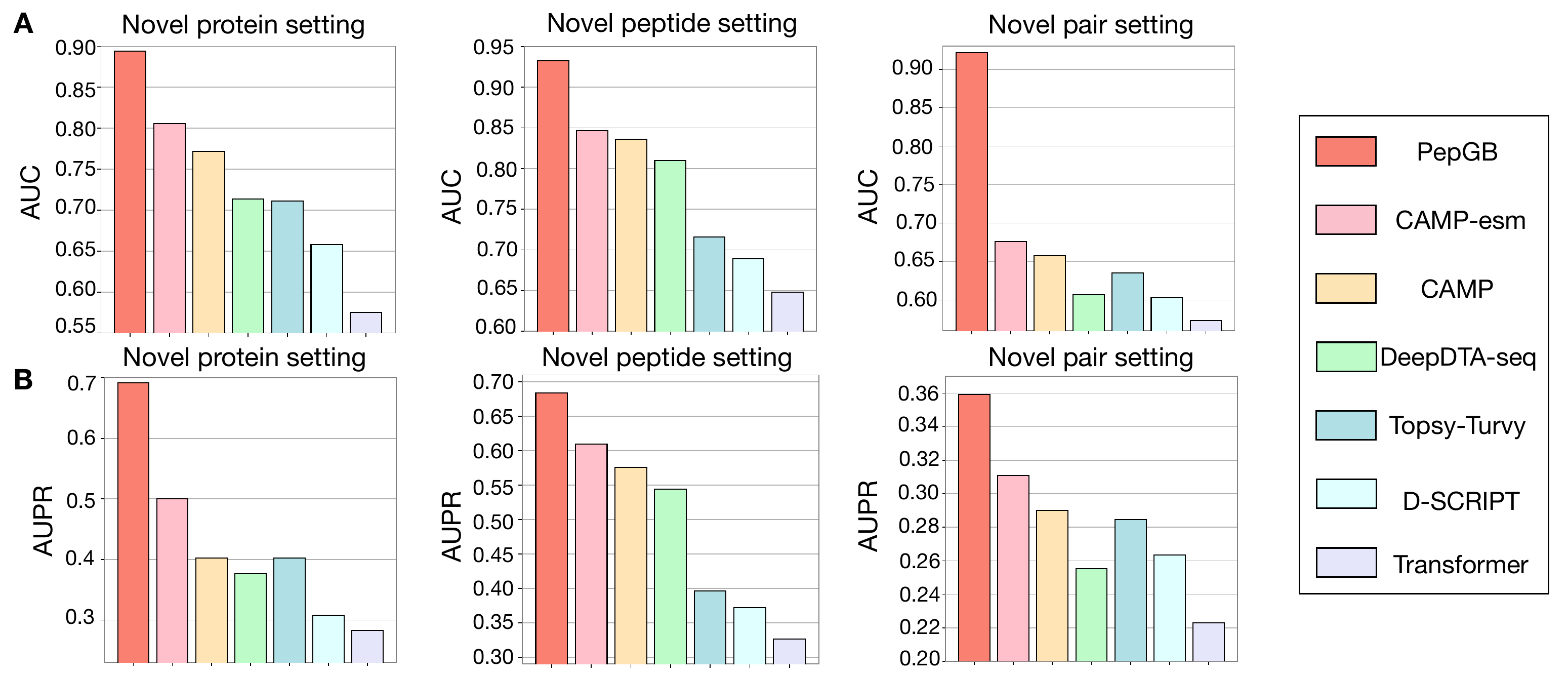}
	\caption{
        Performance of PepGB and other baselines on binary PepPI prediction. \textbf{A} and \textbf{B} show the AUC and AUPR scores of PepGB and six baseline methods under three cross-validation settings, respectively.
     }
	\label{fig:fig4}
\end{figure}

To systematically evaluate the performance of PepGB, we carefully compared our model with six state-of-the-art methods under three rigorous cross-validation settings on the binary interaction benchmark. To ensure a fair comparison, all baseline models were retrained on our benchmark dataset through cross-validation. Model performance was evaluated using the area under the receiver operating characteristics curve (AUC) and the area under the precision-recall curve (AUPR), with reported averages from 5 repeats. Detailed statistics of mean and standard deviation are provided in Supplementary Tables S1 and S2. As shown in Fig. \ref{fig:fig4}A and B, PepGB consistently outperformed all baselines, with an increase by at least 9\%, 9\% and 27\% in terms of AUC score and an increase by at least 19\%, 6\% and 4\% in terms of AUPR score under the ``novel protein setting'', ``novel peptide setting'' and ``novel pair setting'', respectively. Interestingly, under the ``novel pair setting'', we observed the AUC score almost remained stable while the AUPR score exhibited a significant decrease. This indicated that although the AUC margin loss helped PepGB maintain relatively high AUC scores, the prediction performance on positive edges actually dropped due to the difficulty setting. Overall, the results showed that PepGB possessed superior performance in predicting PepPIs for novel peptides and proteins.

\begin{figure}[htbp]
	\centering
	\includegraphics[width=0.9\textwidth]{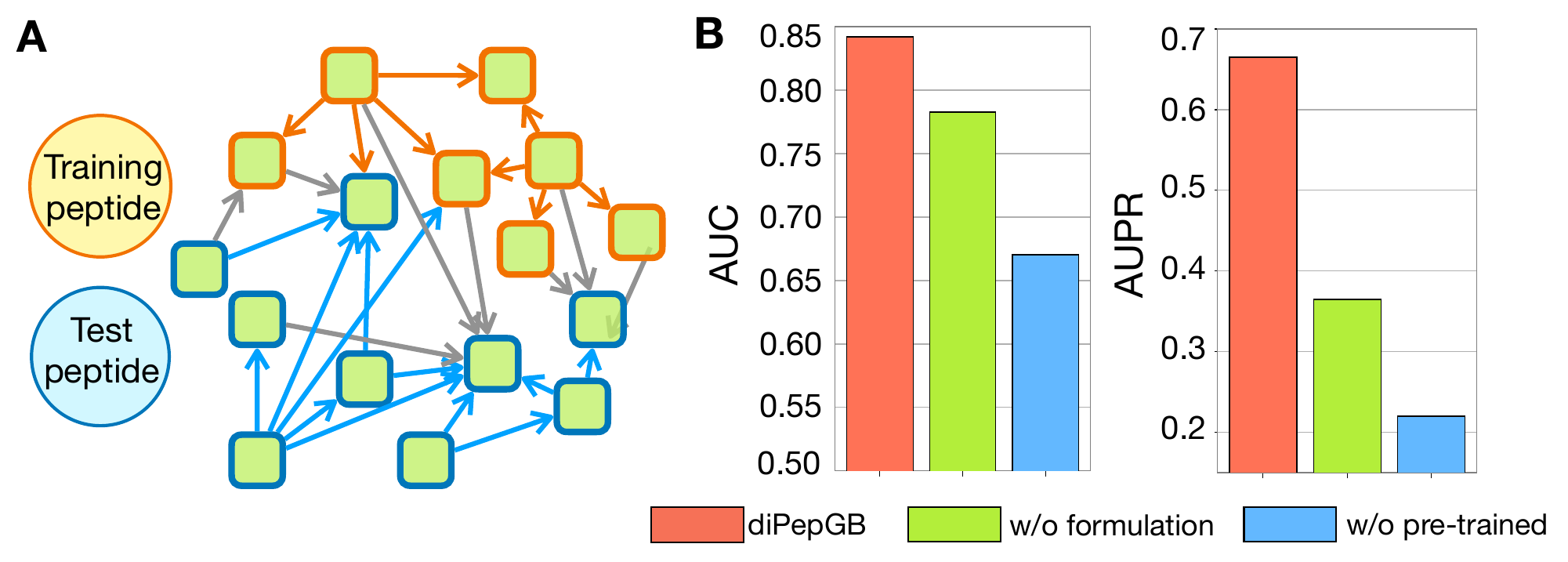}
	\caption{
        Performance of diPepGB on PMI peptide analogs that bind to the anti-tumor target MDM2. \textbf{A} The data splitting strategy and validation setting of diPepGB. \textbf{B} The AUC and AUPR scores of diPepGB and two baselines. More specifically, diPepGB outperformed two baselines both in terms of AUC and AUPR scores.
     }
	\label{fig:fig5}
\end{figure}

\subsection{diPepGB overcomes the uneven issue in modeling highly imbalanced data for lead generation and optimization processes}\label{sec:diPepGB}

Highly imbalanced data are prevalent in assays generated during the lead generation and optimization stages. Nevertheless, constructing the PepPI graph based on extremely imbalanced data results in an uneven local topological structure (Fig. \ref{fig:fig2}A) where peptide nodes are almost isolated without directly mutual message passing. To solve this issue, we introduced diPepGB, by augmenting directed edges between peptide nodes from the same assay. For diPepGB, we compared each pair of peptides within the same assay and constructed directed edges if the binding affinity of the source peptide node was significantly stronger (at least threefold) than that of the destination node.

To demonstrate the advantage of diPepGB, we utilized PepPI data from two assays targeting the same onconprotein MDM2. MDM2 plays a vital role in cancer therapy by negatively regulating the tumor suppressor protein p53 \cite{pmi1}. Previous studies have identified a potent peptide inhibitor, PMI (TSFAEYWNLLSP) of the p53-MDM2 interactions. Consequently, researchers explored multiple PMI peptide analogs by altering residues for lead optimization, providing useful data for our evaluation.

To assess the prediction capacity of diPepGB, we utilized the binding data from a mutation assay \cite{pmi2} that contains 100 peptides (Fig. \ref{fig:fig2}B).  We first partitioned 80\% of the analogs for training and the rest 20\% for validation. Then we constructed directed edges within the training set and validation set, respectively (Fig. \ref{fig:fig5}A), ensuring no overlapped peptides or linked edges between the training and validation set. 

We introduced two baselines for comparison: a regression model based on CAMP architecture for binding affinity prediction (denoted as ``w/o formulation'') to show task difficulty by a conventional regression paradigm and a model with randomly initialized features (denoted as ``w/o pre-trained'') to show the importance of pre-trained features. Details of calculating AUC and AUPR scores based on the predicted affinity values can be found in Supplementary S3.2. We observed from Fig. \ref{fig:fig5}B that, diPepGB greatly outperformed other methods with an AUC score of 0.842 and an AUPR score of 0.665. This substantial improvement underscored the significant contributions of the rank-based graph formulation and pre-trained features to diPepGB. In summary, our results highlighted diPepGB's ability to predict relative binding strength in novel peptides, offering potential applications in lead generation and optimization.

\begin{figure*}[!t]%
\centering
\includegraphics[width=0.9\textwidth]{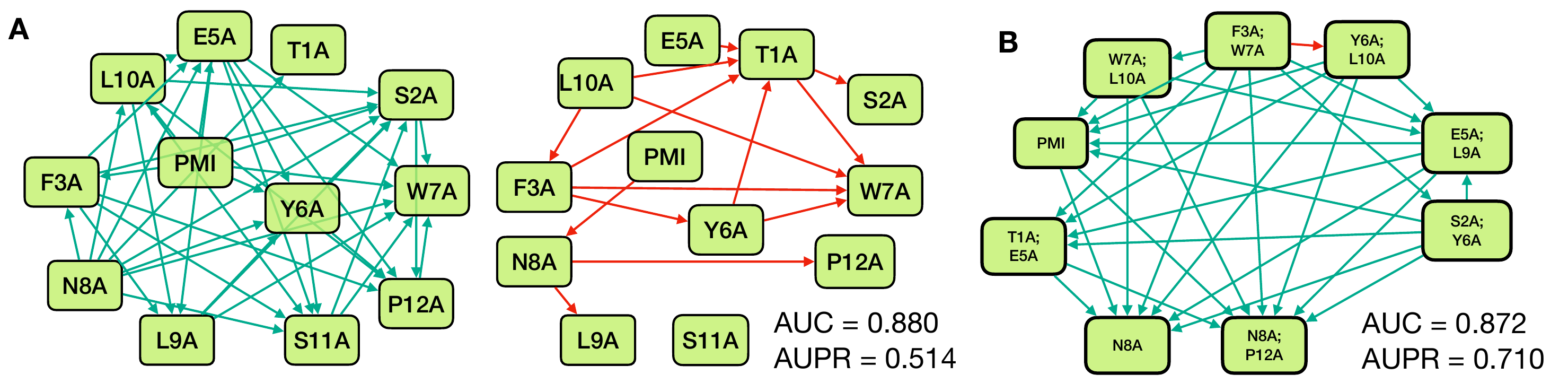}
\caption{Performance of diPepGB on two alanine scanning assays. The directed edges that were correctly predicted by diPepGB were colored in green and edges that diPepGB failed to retrieve were colored in red.  \textbf{A} The case study of a single-substitution alanine scanning assay.\textbf{B} The case study of a two-substitution alanine scanning assay. }\label{fig:fig6}
\end{figure*}

\subsection{diPepGB can be applied for virtual alanine scanning}\label{sec:diPepGB_ala}
The alanine screening assay, a standard strategy to assess the contribution of individual residues to the overall binding activities, involves substituting each peptide residue with alanine iteratively for affinity measurement. Widely applied in lead generation and optimization, we explored the applicability of diPepGB in virtual alanine scanning. diPepGB is trained on complete data from the aforementioned assay in Sec. \ref{sec:diPepGB}, and two independent test sets were collected: a single-substitution alanine screening assay with 12 PMI mutants \cite{pmi3} and a two-substitution alanine screening assay with 9 PMI mutants \cite{pmi4}. Pairwise comparisons on mutants within each testing assay are conducted to construct directed edges, with five negative edges sampled for each positive. To intuitively demonstrate the accurate prediction, we visualized the prediction results by plotting directed edges that were correctly predicted by diPepGB and ones that diPepGB failed to retrieve, respectively. Considering the imbalanced training data, the 75\% quantile of predicted scores was chosen as the threshold to characterize positives and negatives. Fig. \ref{fig:fig6}A and B showed that diPepGB achieved satisfying generalizability with an AUC score of 0.880 and an AUPR score of 0.514 on the single-substitution assay (Fig. \ref{fig:fig6}A), an AUC score of 0.872 and an AUPR score of 0.710 on the two-substitution assay (Fig. \ref{fig:fig6}B), respectively. The results illustrated that diPepGB recapitulated most topological information, indicating its ability to help researchers in deciphering residue-level binding activities through virtual alanine scanning.

\subsection{Analysis of critical model design of PepGB}\label{ablation}
Having validated that PepGB achieved outstanding performance for PepPI. We first investigated the impact of different graph neural network architectures by two variants: PepGB-GraphSAGE and PepGB-GIN. Using GraphSAGE \cite{graphsage} and GIN \cite{gin}, these two variants differ in their aggregation mechanisms. GraphSAGE \cite{graphsage} aggregates information from neighbors and takes the mean to update node features. Graph isomorphism network (GIN) \cite{gin} exploits multi-layer perceptrons for aggregating neighboring information. In addition, two variants, ``PepGB-no DropMessage" (removing the DropMessage module) and ``PepGB-BCE" (only using standard binary cross-entropy loss), are introduced to assess the importance of the DropMessage module and the dual-view loss. We compared the performance of PepGB against these variants under the ``novel protein setting'' and reported the mean and standard deviation over five repeats to obtain robust results. We observed from Table \ref{tab:abl_study} that, PepGB largely surpassed two graph-based variants in terms of AUC and AUPR scores, demonstrating that incorporating GAT into PepGB was more suitable for binary PepPI prediction task. Furthermore, PepGB outperformed ``PepGB-no DropMessage'' and ``PepGB-BCE'', emphasizing the importance of the two modules. Overall, the above results  and effectiveness of each designed component.

\begin{table}[!t]
\caption{The results of ablation studies on PepGB. The mean and standard deviation of five repeats are shown.}\label{tab1}%
\begin{tabular*}{\columnwidth}{@{\extracolsep\fill}lll@{\extracolsep\fill}}
\toprule
   & AUC  &  AUPR \\
\midrule
PepGB-GraphSAGE   &  0.840 $\pm$ 0.026 & 0.518 $\pm$ 0.049\\
PepGB-GIN   & 0.793 $\pm$ 0.036  &  0.423  $\pm$ 0.081\\
PepGB-no DropMessage & 0.862 $\pm$ 0.025 & 0.438 $\pm$ 0.065  \\
PepGB-BCE    & 0.867 $\pm$ 0.021 & 0.557 $\pm$ 0.078\\
PepGB    & 0.896 $\pm$  0.026 & 0.586 $\pm$ 0.087\\
\bottomrule \label{tab:abl_study}
\end{tabular*}
\end{table}

\section{Discussion}\label{sec:discuss}

In this study, we propose PepGB, a graph-based deep learning framework designed to predict peptide-protein interactions to facilitate peptide drug discovery. We first pointed out that the limited size, imbalanced data distribution and inconsistent data quality hinder the application of existing models in peptide early drug discovery. To tackle the challenges, PepGB leverages graph attention neural networks to capture the mutual information within a heterogeneous PepPI graph containing limited peptides and proteins. To avoid overfitting, PepGB integrates the DropMessage module to add fine-grained perturbation during message passing. Besides, PepGB adopts a dual-view loss to improve its robustness on imbalanced and noisy data. Through rigorous evaluation in three settings, we demonstrated the superior performance of PepGB on predicting PepPIs for novel targets and peptide hits. We further derived diPepGB, an extended version of PepGB, to enhance the predicting performance on PepPI graphs with uneven topological structures by comparing pairwise peptide binding affinities to construct error-tolerated directed edges. Evaluations on real-world assays revealed that diPepGB addressed the uneven topological issue and formulated solutions for imbalanced experimental data prone to systematic errors and batch effects, highlighting its potential in applications in lead generation and optimization processes. In general, we believe our work can serve as a valuable and useful tool for the PepPI prediction and thus facilitates peptide early drug discovery process.

\section{Acknowledgements}

This work was supported by the National Natural Science Foundation of China (T2125007 to J.Z.), the National Key Research and Development Program of China (2021YFF1201300 to J.Z.), the New Cornerstone Science Foundation through the XPLORER PRIZE (J.Z.), the Research Center for Industries of the Future (RCIF) at Westlake University (J.Z.), and the Westlake Education Foundation (J.Z.).

\clearpage

\setcounter{equation}{0}
\renewcommand\theequation{S.\arabic{equation}}
\setcounter{table}{0}
\renewcommand{\thetable}{S\arabic{table}}
\setcounter{section}{0}
\renewcommand{\thesection}{S\arabic{section}}

\section*{Supplementary}\label{sec:sl_dataset}

\section{Datasets}\label{sec:sl_dataset}
\subsection{Binary interaction benchmark}
We exclusively utilized interaction data sourced from peptide-protein complexes in the RCSB PDB\cite{pdb1,pdb2} aiming to construct a reliable interaction graph based on explicit physical binding information. Additionally, we augmented the derived PepPI graph with known protein-protein interactions (PPIs) supported by clear physical binding evidence. We found 5,902 processed protein-protein complexes from a previous work \cite{masif} and after mapping by their UniProt ids \cite{uniprot1,uniprot2}, we obtain 191 overlapped PPI edges (defined as both proteins of the PPI exist in PepPI graph). Furthermore, we we incorporated 2,139 overlapping PPI edges with positive experimental scores or from the experimental system ``Co-crystal Structure'' from a protein-protein interaction database BioGRID \cite{biogrid1,biogrid2}. Finally, we incorporated these physical PPI edges into our PepPI graph only for message passing and these PPI edges were not involved in supervised link prediction.

\subsection{Peptide mutation data}
We first collected 100 PMI mutants from a previous work \cite{pmi2}, in which researchers substituted residues from nine positions of the wild-type peptide and obtained over 100 peptide analogs of PMI (TSFAEYWNLLSP) via fluorescence polarization techniques and surface plasmon resonance (SPR). This set includes 22 single-substitution analogs, 5 multi-substitution analogs with corresponding SPR-measured binding affinities, and 73 two-substitution analogs without observed binding activities. Pairwise comparisons between the 27 positives and 73 negatives were conducted, assigning directed edges from stronger peptides to weaker ones when the binding affinity difference exceeded threefold. Negatives were assigned an extremely large Kd value of 100,000 nM, and no comparisons were made among negatives.

Next, we collected 12 single-substitution analogs and 9 two-substitution analogs from two alanine scanning mutation analysis \cite{pmi3,pmi4}, respectively. Based on the 12-mer PMI peptide (TSFAEYWNLLSP), researchers iteratively substituted one or two residues by alanines at each position. The binding affinities of PMI and Ala-substituted analogs were measured using the SPR-based competition binding assay. In the single-substitution assay, we conducted pairwise comparison on binding affinities and successfully constructed 54 directed edges among 132 node pairs. In the two-substitution analogs, we conducted pairwise comparison on binding affinities and successfully constructed 31 directed edges among 72 node pairs.

\section{Validation settings}\label{sec:sl_validation}
\subsection{Similarity-based clustering}
Here the similarity between two amino acid sequences $v_i$ and $v_j$ is defined as
\begin{equation}
\frac{\text{SW}(v_i,v_j)}{\sqrt{(\text{SW}(v_i,v_i)\text{SW}(v_j,v_j))}},
\end{equation}\label{eq:sim}
where SW($\cdot,\cdot$) represents the Smith-Waterman alignment score (\url{https://github.com/mengyao/Complete-Striped-Smith-Waterman-Library)}) between two sequences. Then we applied a single-linkage clustering algorithm and limited the maximal similarities by a pre-defined threshold between any two mode from different clusters. 

The similarity threshold should achieve a delicate balance: large enough to distinguish sequences between training and testing datasets, but can not be too large since it may cause a less clusters with extremely large cluster sizes and thus influence the data splitting process cross-validation. On the other hand, a threshold that is too small may yield nearly random splitting results. In alignment with previous studies \cite{monn,camp}, we balanced the trade-off and picked similarity thresholds of $0.4$ for peptides and $0.5$ for proteins in our benchmark dataset. Clustered by such thresholds, the sequences within each cluster was approximately evenly distributed.

\subsection{Cross-validation setting}
We conducted five-fold cross-validation on the sequence clusters for the ``novel peptide setting'' as well as the ``novel protein setting'' and the proportion of validation set was roughly 20\%. For the ``novel pair setting'', we intricately partitioned the peptide clusters into three grids and further subdivided the protein clusters into three grids within each peptide grid. In such a manner, we approximately divided the dataset into nine grids for nine-fold cross-validation. We used the PepPI edges from one grid as the validation set and the rest four grids that did not share any protein or peptide clusters for training. This approach ensured the absence of similar peptide or protein nodes across the training and testing sets.

\section{Baselines}\label{sec:sl_baseline }
\subsection{Baseline methods and metrics of PepGB}
We compared PepGB with several existing deep learning methods under three validation settings.  To adapt DeepDTA-seq \cite{deepdta} for peptide-protein interaction (PepPI) classification, we employed a modified version that incorporated learnable word embedding features for peptides and proteins, utilizing a binary prediction head with a sigmoid function. In particular, for our previous work CAMP and CAMP-esm \cite{camp}, we adopted the default parameters from the original paper. For DeepDTA-seq \cite{deepdta}, we conducted a grid search to determine the best combination of hyper-parameters, including the length of sequence window from [4,6,8,12], and we used 100 as the maximum number of epochs, which was the default value from the original paper. For D-script \cite{d_script} and its follow-up work Topsy-Turvy \cite{tt}, we conducted the same grid search scheme to determine the best combination of hyper-parameters, including the batch size from [32,64,128,256], learning rate from [0.1,0.05,0.01,0.005,0.001,0.0005,0.0001], lambda from [0.2,0.4,0.6,0.8] and number of epochs from [25,50,100]. We also added the hyper-parameter 'no-augment' since the peptide-protein pairs can not be reversed for data augmentation. For the Transformer model \cite{attention}, we used a learnable word embedding layer to embed each amino acid of the protein or peptide sequences into a 128-dimensional vector. We used peptide sequences as the``query", and protein sequences as the ``key" and ``value" in the attention module of the Transformer. The hyper-parameters in our search scheme included combinations of batch sizes from [32, 64, 128] and learning rates from [0.0005, 0.0001, 0.00005]. To compare fairly, all baseline methods were trained and evaluated using the same settings as PepGB. The only difference is that, for these baselines, we shuffled all non-interacted peptide-protein pairs to generate ``negatives'' in advanced while PepGB randomly samples non-existing negative edges at each training epoch. 

As mentioned in the main text, we also tried, PepNN \cite{pepnn},  a deep learning method to identify peptide binding residues from protein surface. PepNN takes the peptide sequence and protein sequence or structure as input so we speculated this framework could be transfer to predict binary peptide-protein interactions.  We first tried to replace its output head (originally using a softmax layer to generate binding scores for individual protein residues) with a binary prediction head (a max pooling layer plus fully connect layers with a sigmoid function) and re-trained PepNN with PepPI data. We also tried to directly use PepNN for inference by calculating the average or maximum binding score over all protein residues. These attempts only yielded prediction AUC scores oscillating around 0.5, indicating unsuitability for this task.

We also attempted to benchmark PepGB with some structure-based methods to estimate how the huge conformation flexibility of peptides would influence the prediction result. More specifically, we downloaded the crystal structures of peptide-protein complexes in the benchmark dataset and retrieved the 3D structures of the peptide chains and protein chains, respectively. We tried ProNet \cite{pronet}, a 3D graph framework that hierarchically represents the protein or peptide structure at residue level, backbone level and all atom level. However, when we trained ProNet on our peptide-protein data, the validation AUC remained fluctuated around 0.5. And same condition recurred when we used GearNet \cite{gearnet}, a structure-based framework that encodes the protein or peptide structures into residue-level 3D graphs. One possible reason might be that these struture-based methods represent peptide structures as a residue-level geometric graph, thus fail to capture the extensive conformation flexibility of peptide structures. 
Although our expertise in structural modeling was limited, these initial attempts suggest the need for tailored frameworks for modeling peptide-protein interactions from a structural perspective in the future.

To evaluate the prediction performance of PepGB and other methods, we chose the area under the receiver operating characteristics curve (AUC) and the area under the precision-recall curve (AUPR) as metrics.

\subsection{Baseline methods and metrics of diPepGB}
In real-world biological field, there exist extensive binding assays with extremely imbalanced data distribution. One common example is the mutation analysis, where the binding affinities of a series of peptide analogs targeting the same protein are measured. We therefore propose diPepGB to address the bottleneck of modeling such extremely uneven data via graphs. We compared diPepGB with two baselines. First, we constructed a regression model as a conventional paradigm by adopting the model architecture of CAMP \cite{camp} with a regression head, denoted as ``w/o formulation'', to directly model the binding affinities of the mutation dataset and observed an overall spearman correlation 0.5608. We further made pairwise comparison based on the predicted affinities to construct ``predicted directed edges''. We then calculated the AUC and AUPR scores between these predicted directed edges and true edges. Since we only constructed directed edges when the source peptide is significantly stronger than the destination peptide, thus can alleviate the influence of systemic error to a certain degree.
Furthermore, to evaluate the contribution of pre-trained peptide features, we only used a directed graph with the same GNN architecture and replaced the pre-trained node features by random initialized vectors. Hyper-parameters of diPepGB are used for these two methods for consistency.

\section{Training details}
We utilized a contrastive learning-based pre-trained sequence encoder to extract peptide features and we directly used the pre-trained protein language model ESM2 \cite{esm} (\url{esm2_t33_650M_UR50D}) to extract protein features. Then PepGB averages the embeddings along the sequence dimension as individual node feature ($d=1280$). 

PepGB consists of two graph attention neural layers of hidden size 512 and a DropMessage module with dropout rate $p=0.5$. We set learning rate to be $10^{-4}$ and used Adam optimizer with the decay rate of the first and second moments $\beta_1 = 0.9, \beta_2 = 0.999$, respectively. The training process contained 50 epochs with an early-stopped mechanism in terms of validation AUC scores. For each epoch, the disjoint train ratio is set to 0.4, which indicates that 40\% of the edges are used for supervised learning 60\% of the edges are used only for message passing. Upon the AUC min-max margin loss, we applied the default value of margin $m=1$, the weight of the binary cross-entropy loss is $\eta=0.3$ and the weight of AUC min-max margin loss is $0.7$. For the pre-training stage, the batch size is set to be 128 and temperature $\tau$ in the InfoNCE loss is set to be 0.05. diPepGB inherits the above hyper-parameters with an additional self-loop edges to maintain feature information about a node itself. All these hyper-parameters are determined using a grid search approach. To facilitate the information aggregation and feature updates during training, we enable message passing through all nodes on the complete graph.

\begin{table}[t]
\caption{AUC of PepGB and other baselines for PepPI prediction under three evaluation settings. The mean and standard deviation of five repeats are reported.}\label{tab:tabs1}%
\begin{tabular*}{\columnwidth}{@{\extracolsep\fill}llll@{\extracolsep\fill}}
\toprule
   & Novel protein  &  Novel peptide   &  Novel pair \\
\midrule
PepGB   & 0.8942 $\pm$ 0.0300   & 0.9326 $\pm$ 0.0362  & 0.9215 $\pm$ 0.0137\\
CAMP   & 0.7715 $\pm$ 0.0235  & 0.8359 $\pm$ 0.0523 & 0.6578 $\pm$ 0.0141\\
CAMP-ESM   & 0.8058 $\pm$ 0.0091   & 0.8468 $\pm$ 0.0506  & 0.6762 $\pm$ 0.0189\\
D-script   & 0.6581  $\pm$ 0.0190  & 0.6891 $\pm$ 0.0285  &  0.6029 $\pm$ 0.0183\\
Topsy-Turvy    & 0.7110 $\pm$ 0.0219  & 0.7158 $\pm$ 0.0335  & 0.6325 $\pm$ 0.0169\\
DeepDTA-seq    & 0.7133 $\pm$ 0.0325 & 0.8097 $\pm$ 0.0569  & 0.6071 $\pm$ 0.0091\\
Transformer    & 0.5751 $\pm$ 0.0108  & 0.6482 $\pm$ 0.0076  & 0.5731 $\pm$ 0.0175\\
\bottomrule
\end{tabular*}
\end{table}

\begin{table}[t]
\caption{AUPR of PepGB and other baselines for PepPI prediction under three evaluation settings. The mean and standard deviation of five repeats are reported.}\label{tab:tabs2}%
\begin{tabular*}{\columnwidth}{@{\extracolsep\fill}llll@{\extracolsep\fill}}
\toprule
   & Novel protein  &  Novel peptide   &  Novel pair \\
\midrule
PepGB   & 0.6916 $\pm$ 0.1158   & 0.6651 $\pm$ 0.1158 & 0.3532 $\pm$ 0.0404 \\
CAMP   & 0.4424  $\pm$ 0.0172  & 0.5754 $\pm$ 0.0798 & 0.2830 $\pm$ 0.0221\\
CAMP-ESM   & 0.5005 $\pm$ 0.0211  & 0.6092 $\pm$ 0.0748 & 0.3108 $\pm$ 0.0188\\
D-script   & 0.3079 $\pm$ 0.0249  & 0.3718 $\pm$ 0.0307 & 0.2634 $\pm$ 0.0209\\
Topsy-Turvy    & 0.4023 $\pm$ 0.0319 & 0.3964 $\pm$ 0.0463  & 0.2822 $\pm$ 0.0214\\
DeepDTA-seq    & 0.3764 $\pm$ 0.0410 & 0.5441 $\pm$ 0.0841  & 0.2552 $\pm$ 0.0113\\
Transformer    & 0.2831 $\pm$ 0.0069 & 0.3261 $\pm$ 0.0090  & 0.2229 $\pm$ 0.01127\\
\bottomrule
\end{tabular*}
\end{table}

\clearpage
\bibliographystyle{unsrt}
\bibliography{ref_main}

\end{document}